\documentclass[aps,pre,twocolumn,showpacs,endfloats*]{revtex4}
\usepackage{natbib}
\usepackage{epsfig}
\usepackage{amsmath}

\def\invitro{\emph{in vitro}}
\def\invivo{\emph{in vivo}}

\def\ds{\partial_s}

\def\pN{\,\mbox{pN}}
\def\mum{\,\mu\mbox{m}}
\def\nm{\,\mbox{nm}}
\def\s{\,\mbox{s}}
\def\muM{\,\mu\mbox{M}}
\def\cos{\mbox{cos}\,}
\def\sin{\mbox{sin}\,}
\def\tan{\mbox{tan}\,}

\def\ecoli{\emph{E. coli}}
\def\bsub{\emph{B. subtilis}}
\def\caul{\emph{C. crescentus}}

\def\therm{\emph{Thermotoga}}

\def\kbon{k^b_{\mbox{\scriptsize on}}}
\def\kboff{k^b_{\mbox{\scriptsize off}}}
\def\kpon{k^p_{\mbox{\scriptsize on}}}
\def\kpoff{k^p_{\mbox{\scriptsize off}}}
\def\kon{k_{\mbox{\scriptsize on}}}
\def\koff{k_{\mbox{\scriptsize off}}}

\begin{document}
\newcommand{\citel}[1]{\citep{#1}}
\newcommand{\citelt}[1]{\citet{#1}}
\title{Steady-state MreB helices inside bacteria: dynamics without motors}

\author{Jun F. Allard}
\author{Andrew D. Rutenberg}
\email{andrew.rutenberg@dal.ca}
\homepage{http://www.physics.dal.ca/~adr}
\affiliation{Department of Physics and Atmospheric Science, 
Dalhousie University, Halifax, Nova Scotia, Canada, B3H 3J5}

\date{\today}

\begin{abstract}
Within individual bacteria, we combine force-dependent polymerization dynamics of 
individual MreB protofilaments with an elastic model of protofilament bundles
buckled into helical configurations.
We use variational techniques and stochastic simulations to relate
the pitch of the MreB helix, the total abundance of MreB, and the 
number of protofilaments.  By comparing our simulations with mean-field 
calculations, we find that stochastic fluctuations are significant.
We examine the quasi-static evolution of the helical pitch with cell growth, as
well as timescales of helix turnover and denovo establishment. 
We find that while the body of a polarized MreB helix treadmills towards its
slow-growing end, the fast-growing tips of laterally associated protofilaments
move towards the opposite fast-growing end of the MreB helix.  This offers a possible 
mechanism for targeted polar localization without cytoplasmic motor proteins.
\end{abstract}
\maketitle

\section{Introduction}

The eukaryotic cytoskeleton organizes cell shape, cell polarity, cell division, 
and non-diffusive subcellular transport. F-actin, microtubules, and intermediate 
filaments comprise the cytoskeleton, and act with the associated proteins that 
provides spatial and dynamic control of cytoskeletal function \citel{mre52}.  
Prokaryotic cells have cytoskeletal analogues, such as the FtsZ-ring
associated with cell division \citel{mre86} together with its ``divisome'' 
of associated proteins.  Bacteria also have a number of polymerizing cytoplasmic proteins, 
such as ParM \cite{mre55} and MinD \cite{mre86} 
that exhibit distinctive helical structures within the cell.

Recently, the actin homologue MreB has been shown to play a cytoskeletal 
role in many bacteria \citel{mre92, mre87, mre86}. MreB forms a continuous cytoplasmic 
helix that runs the length of nearly all rod-shaped prokaryotes, including 
{\em Escherichia coli}, {\em Bacillus subtilis} and {\em Caulobacter crescentus} 
\citel{mre30}, and it has been implicated in shape determination and polar protein localization.

In most Gram-positive bacteria MreB is 
present together with several paralogues, such as Mbl and MreBH in \bsub. 
The helical pitches for MreB or Mbl, separately observed by immunoflourescence microscopy, 
are reported to be $0.73 \pm 0.12 \mum$ and $1.7 \pm 0.28 \mum$, respectively 
\citel{mre13}. More recent measurements of fluorescent fusions of MreB and of Mbl 
report pitches of $0.6 \pm 0.14 \mum$, with colocalization \citel{mre90}. 
In Gram-negative species, such as {\em E. coli} and {\em C. crescentus},
only MreB is present. In \ecoli, pitches of $0.46 \pm 0.08 \mum$ have 
been reported \citel{mre73}. In all cases, the helices are dynamic, with elements
moving along the main helix at reported speeds ranging from 
$6 \nm/\s$ \citel{mre71} to $70 \nm/\s$ \citel{mre25}. The helical structure has also
been observed to condense into a ring at midcell near the time of division in 
\ecoli\ \citel{mre45}, \caul\ \citel{mre23} and \bsub\ (where only MreBH 
coils) \citel{mre88}.

The MreB helix appears to be composed of a bundle of individual 
``protofilaments'' \citel{mre29,mre33,protofilament}. Quantitative 
immunoblotting has been used to measure the molecular abundance of various 
MreBs. In \bsub, there are roughly 8000 MreB monomers and $12000-14000$ monomers of 
Mbl \citel{mre13} while \ecoli\ has roughly $17000-40000$ monomers of 
MreB \citel{mre73}. Neglecting the cytoplasmic fraction of monomeric MreB,
these abundances suggest a bundle thickness of about 10 protofilaments 
\citel{mre39}. 

In \bsub\, Mbl is necessary for proper insertion of new peptidoglycan, which occurs in a 
helical fashion \citel{mre30}, while MreBH is necessary for the localization and function of the 
cell wall hydrolase LytE that is believed to recycle the outer layers of the cell wall, also in a 
helical fashion \citel{mre88}. Cells with mutant {\em mreB} are wide, rounded and usually not 
viable \citel{mre67}.  Helical bundles of MreB may contribute to the spatial localization
of associated MreC, MreD, and PBP2 that, in turn, help to coordinate cell wall synthesis. 
It has also been suggested that helical filaments of MreB paralogues under tension 
can lead to spiral morphologies \cite{mre65}.

Disruption of MreB leads to loss of proper polar 
localization of a number of proteins such as the chemotaxis protein
Tar and the {\em Shigella flexneri} virulence factor IcsA in 
\ecoli\ \citel{mre37}, and three integral membrane proteins (PleC, DivJ, CckA) in \caul\ 
\citel{mre34}.  Polar localization in \caul\ was disrupted by either 
underexpression or overexpression of MreB.  When normal MreB expression was
returned, polar localization was re-established \citel{mre34}.
This suggests that MreB has a continual role in either direct polar trafficking of 
these proteins or in the maintenance of landmarks necessary for their proper 
positioning \citel{mre92}.   

The polar proteins in \caul\ are normally directed towards distinct (stalked and swarmer) 
poles in different stages (swarmer, stalked, and predivisional) of its life cycle. For example, 
PleC is localized to swarmer poles in swarmer and predivisional cells, DivJ is localized to 
stalked poles, while CckA is localized to both poles of predivisional cells \citel{mre50}. After 
MreB expression is disrupted and restored, PleC and DivJ are restored {\em randomly} to either pole 
\citel{mre50}.  This suggests that MreB may be polarized within \caul\ and that the polarity of the 
MreB helix is randomly restored after its disruption.  However, tracking of individual YFP-MreB 
molecules shows unpolarized motion \citel{mre71}, raising questions about the mechanism of 
specific polar targeting. MreB-directed targeting to specific poles 
has not been reported in \ecoli\ or \bsub.

MreB interacts with both RNA polymerase (RNAP) \citel{mre50} and SetB, a 
chromosome defect suppressor \citel{mre53}, and has been implicated in 
the fast polar translocation of the origin-proximal regions ({\em oriC}) \cite{mre91} of newly-replicated 
DNA in \caul\ \cite{mre26} and in \ecoli\ \cite{mre50} (see however \cite{blaauwen}).  
Time-lapse microscopy has shown that the polar transport of {\em oriC} 
in \bsub\ had an average speed of $2.8 \nm/s$ and a peak speed of $4.5 \nm/s$ \citel{mre91}. 

MreB is a homologue of the eukaryotic cytoskeletal protein actin
\citel{act04,act05}. Actin filaments are used to change
cell shape and to move bacteria such as {\em Listeria
monocytogenes} via polymerization forces \cite{act01}, 
and in muscle contraction and for organelle movement via collections of 
myosin motor proteins \citel{mre52}.  Actin assembly is regulated
through a number of ``actin-binding proteins'' that variously 
control cross-linking, bundling, filament nucleation, end-capping, 
filament cutting, monomer sequestration and desequestration. MreB, in contrast, 
does not have any clearly identified motor proteins or associated 
proteins that regulate polymerization.  Notably, MreB filaments 
spontaneously bundle {\em in vitro} without associated proteins 
\citel{mre38}. 
 
The varied roles of MreB inside bacterial cells raise some basic 
questions. What is the origin of its helical configuration, and how significant 
are the forces that the MreB helix applies?  What 
does the helical pitch of MreB filaments depend on? 
What aspects of the MreB system can be 
understood in terms of actin-like polymerization dynamics? 
Specifically, must we invoke yet-to-be discovered prokaryotic 
motor proteins or accessory proteins controlling MreB polymerization 
to explain MreB-related polar localization of proteins such as 
Tar, IcsA, DivJ, PleC, and CckA? 
Finally, does the small size of the bacterial cell affect MreB 
polymerization, as compared to actin polymerization in much larger 
eukaryotic cells?

To address these questions, we present a model of the MreB helix with 
stochastic polymerization dynamics of protofilaments together with 
global elasticity of a helical MreB bundle constrained by the 
bacterial cell wall.  Our model provides a quantitative relationship between helical pitch, 
total abundance of MreB protein in a particular cell and the thickness of the 
protofilament bundles. 
The bundled MreB protofilaments are in a dynamical steady-state, and 
undergo constant advection as the polymerized subunits treadmill.  We 
discuss how this advection could be harnessed for targeted polar localization
without motor proteins.  The steady-state dynamics also allow us to address other dynamical processes
such as protein turnover in FRAP experiments \citel{mre28}
or  recovery from A22-induced disruptions of the MreB helix \citel{mre34}. 

\section{MODEL}

\subsection{Protofilament polymerization}

Both actin and MreB polymerize into polarized filaments. Addition and dissociation of 
monomers occur at the ends of the asymmetric filament, both at the 
``barbed" (``$+$'' or fast-growing) tip and the ``pointed" (``$-$'' or slow-growing) 
tip. The kinetics of actin polymerization are well-characterized by 
concentration-dependent polymerization rates $\kbon c$ and $\kpon c$ at 
the barbed and pointed ends, respectively, where $c$ is the cytoplasmic 
monomer concentration, and concentration-independent depolymerization rates $\kboff $ and $\kpoff $.

If a force $F$ is applied to a filament's tip, the polymerization rate is 
reduced. When thermal bending fluctuations are much faster 
than the polymerization dynamics, the polymerization rates at either end of the 
filament are reduced to 
\citel{act01}
\begin{equation} 
 	\kon c e^{-{F a_0}/{k_B T}},
	\label{eq::konf}
\end{equation}
where $a_0 = 5.1 \nm$ is the MreB monomer length \citel{mre72} and 
$k_B T = 4.1 \pN \nm$ at room 
temperature.  This force-dependent polymerization rate can also be obtained from 
thermodynamic arguments in the high-force limit \citel{act20}. We apply it to 
MreB polymerization dynamics within the cell.

In the absence of force, each filament grows above, and shrinks below, 
the critical cytoplasmic concentration
\begin{equation}
 	c_c = \frac{\kboff+\kpoff}{\kbon+\kpon}.
	\label{eq::ccdef}
\end{equation}
The asymmetry between the polymerization and depolymerization rates at 
barbed and pointed ends leads to treadmilling, in 
which the filament length remains constant while depolymerization from the 
pointed end is balanced by polymerization at the barbed end \citel{mre52}. 
The force-independent treadmilling rate is \citel{act19}
\begin{equation} \label{eq::lambdatreaddef}
 	\lambda_{tread} = \frac{\kpoff\kbon - \kboff\kpon}{\kpon + \kbon}.
\end{equation}
If the filament position is fixed, treadmilling results 
in a net advection of all polymerized monomers towards the pointed end.

The kinetic rate constants for MreB polymerization are yet to be 
determined explicitly, but appear to differ significantly from eukaryotic 
actin. In addition to spontaneously nucleating and bundling with much greater 
ease than actin, 
purified MreB from {\em Thermotoga maritima} polymerizes \invitro\ faster 
and exhibits a much lower critical concentration ($c_{c, \mbox{\small 
MreB}} = 0.003 \muM$ \cite{mre38} compared with $c_{c, \mbox{\small actin}} = 0.17 
\muM$ \citel{mre71}).  Nevertheless, \invivo\ observations of single 
molecule motion of MreB in \caul\ suggest that the treadmilling rate is 
similar to actin \citel{mre71}.  By starting with \invitro\ rate constants 
from eukaryotic actin \citel{act05} and scaling the on-rates by a factor 
of $c_{c, \mbox{\small actin}} / c_{c, \mbox{\small MreB}} = 55$, we 
recover the reported critical concentration of \invitro\ MreB from  {\em T. maritima}. Following \invitro\
observations that the treadmilling rate is $\lambda_{\mbox{\small MreB}} = 1.2 \s^{-1}$ \citel{mre71}, 
we further scale all four rates by $\lambda_{\mbox{\small MreB}}/\lambda_{\mbox{\small actin}} = 2.1$  --- preserving
the MreB treadmilling rate. These scalings preserve the pointed/barbed-end 
asymmetries of actin and represent the least-intrusive 
modification of actin polymerization dynamics to make it consistent with 
observed MreB dynamics.  The resulting barbed-end polymerization rate constant, $k^b_{on}$ (see Table~1), is close 
to the diffusion limit \cite{mre98} indicating that the pointed/barbed-end asymmetries
of MreB may differ significantly from actin and/or that the appropriate $c_c$ for {\em in vivo} measurements
may be significantly above the {\em T. maritima} value. However, our qualitative results do not depend on the
precise parameter values used in this paper. 

We model polymerization dynamics by the stochastic  
addition/dissociation of monomers at the barbed/pointed ends of individual
MreB protofilaments  \citel{protofilament} using the scaled kinetic rate constants 
discussed above and listed, together with other parameters, in Table~1.  
Except for growing cells in Sec.~\ref{growth}, a standard cell length of 
$L_c=3 \mu m$ and cell radius of $R_c=400 \nm$ are used. 

\subsection{Bundle ultrastructure}

\begin{figure}[h]
\includegraphics[width=8.6cm]{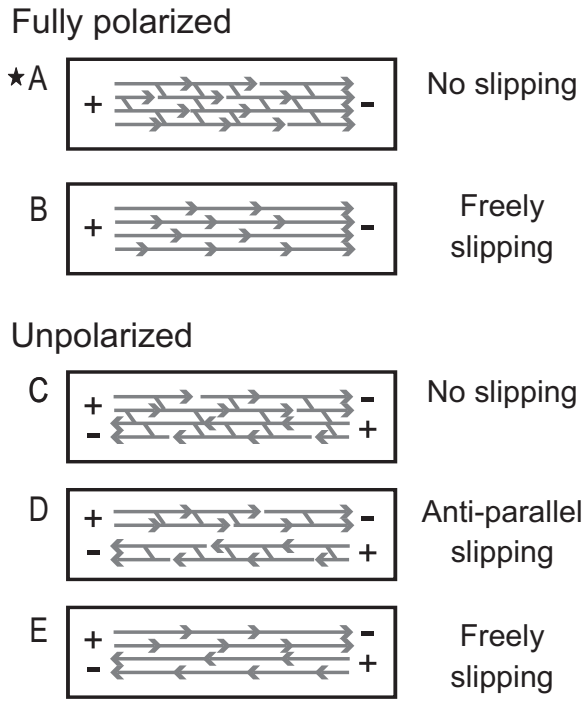}
\caption{Five models for the MreB ultrastructure, each composed of polarized protofilaments. 
A) Single polarized bundle of protofilaments with lateral interactions that prevent 
relative slipping within the bundle. 
As indicated by the star, data in subsequent figures are for this model. 
B) Polarized bundle of protofilaments that freely 
slip relative to each other. C) Unpolarized bundle of protofilaments, with no lateral 
slipping.  
D) Two antiparallel bundles of polarized protofilaments that slip relative 
to each other, though protofilaments within a given bundle do not slip. E) Unpolarized 
bundle of protofilaments that all freely slip with respect to each other.}
\label{FIG::ultrastructures}
\end{figure}

The ultrastructure of the MreB helix -- the precise arrangement, orientation and 
length distribution of the individual protofilaments that make up the helical bundle -- 
remains a mystery.  Several hypotheses have been put forward \citel{mre33} and 
Fig.~\ref{FIG::ultrastructures} illustrates five basic possibilities.  
Several of these are less plausible. The slippery arrays in Fig.~\ref{FIG::ultrastructures}E are unlikely to be able 
to support the forces the cytoskeleton must withstand as it pushes against the cell 
wall, and recent biochemical experiments have demonstrated large lateral 
interactions between filaments \citel{mre38}. The ultrastructure
in Fig.~\ref{FIG::ultrastructures}C leads to tapered bundles as 
antiparallel protofilaments, which cannot slide past each other, treadmill 
in opposite directions.  Such tapering is not seen experimentally. 
In this paper we therefore consider ultrastructures composed of polarized bundle(s) of 
protofilaments: either one bundle (Fig.~\ref{FIG::ultrastructures}A) or two antiparallel 
bundles that freely slide with respect to each other (Fig.~\ref{FIG::ultrastructures}D).  
Since the elastic and polymerization properties of the second case follow straight-forwardly 
from the former, we will mostly focus on a single polarized non-slipping filament bundle 
(Fig.~\ref{FIG::ultrastructures}A) and revisit the possibility of antiparallel bundles 
slipping with respect to each other in the final discussion. 

It is quite possible that individual MreB protofilaments do not continuously
extend from one end of the bacterial cell to the other, similar to actin cables
in yeast \citel{act23}.  Indeed, in \caul\ individual protofilaments 
appear to be much shorter than the cell length, only $392 \pm 23$ nm 
on average \citel{mre71}.
For the purposes of our model, the mechanical and end-polymerization 
properties of discontinuous bundles of protofilaments are equivalent 
to bundles of continuous protofilaments.  Systematic heterogeneities in the MreB helix 
thickness have not been reported, but our model does not depend on how the 
cell regulates the average number of protofilaments in a cross-section of the 
filament bundle.  Protofilament association, dissociation and nucleation are thus 
implicitly included in our model.  
Polymerization and depolymerization  away from the filament edges may occur and
do not affect our steady-state results: whatever the distribution of lateral exchange, in steady state 
monomer incorporation and separation are balanced. 
Of course, significant numbers of 
active protofilament tips away from the cell poles can affect the dynamical
timescales that we discuss. 

\subsection{Elastic bundle}

{\em In vitro}, MreB typically polymerizes into straight filaments 
\citel{mre86}.  However, MreB adopts ring-like coiled configurations in 
spherical mutants of normally rod-shaped organisms \citel{mre07}, and 
forms helices in rod-shaped cells \citel{mre29}.  These observations 
suggest that normally straight elastic MreB protofilaments may simply buckle into 
helices inside the cylindrical confinement provided by the relatively 
hard cell wall.

A self-consistent model for the observed MreB helices consists 
of a particular ultrastructure of protofilaments, as in Fig.~\ref{FIG::ultrastructures}, 
buckled into a helical configuration by the cylindrical cell wall. 
A steady state exists where the polymerization force at the tips of the helices is 
balanced by the mechanical forces of the helical configuration. We treat the cell 
as a spherocylinder with total length $L_c + 2 R_c$ and radius 
$R_c$ and total volume $\pi R_c^2 (L_c + 2 R_c)$. The helix is assumed to 
extend throughout the cylindrical part of the cell, but not into the 
hemispherical poles, as indicated by experiment \citel{mre13}. 

Although helical equilibria of elastic filaments have been investigated 
since the 1800's, they remain a contemporary 
topic \citel{ela14}. In the elastic Cosserat model \citel{ela01}, a 
filament is parametrized by its unstretched arclength $s \in [0,L_{fil}]$ 
where $L_{fil}$ is its total unstretched length. The Hamiltonian of such a 
filament is
\begin{multline} \label{eq::hamiltonian}
 	\mathcal{H} = \frac{1}{2}\int_0^{L_{fil}} 
		[ (B\kappa(s)^2 + C\tau(s)^2) \,{n(s) }^2 \\
 	+ E (1-| \partial \vec{r} / \partial s |)^2 \,n(s) ] ds
\end{multline}
where $\vec{r}(s)$ is the position of the centerline, $\kappa(s)$ is the 
local curvature, $\tau(s)$ is the local twist and $n(s)$ is the local 
filament thickness (measured in number of protofilaments). $B$, $C$ and $E$ 
are the bending, twisting and stretching modulii of an individual protofilament, respectively. For an actin bundle, 
the thickness dependence of bending and twisting ranges from linear ($n$) for slippery protofilaments 
to quadratic ($n^2$) for non-sliding protofilaments, with a crosslinker dependent crossover \cite{ela16}.  
For MreB, lateral interactions appear strong so we assume a quadratic dependence on $n$.

By imposing the observed helical configuration we can use variational techniques to 
estimate the forces working against monomer addition at the filament tips. 
In Appendix~\ref{appendix::globalelastic}, 
we derive the force acting at the filament tips along the filament direction in the
inextensible ($E \rightarrow \infty$) and freely twisting ($\tau(s)=0$) regime:
\begin{equation}
 	 F_B = \begin{cases} 
 	f_B \;\mbox{sin}^2 \theta\; (1+ 3\,\mbox{cos}^2\theta )  
			\langle n^2 \rangle & L_{fil} > L_c \\
 	0 & L_c > L_{fil},
  \end{cases}
	\label{eq::efm1}
\end{equation}
where $\theta$ is the pitch angle of the helix and $f_B \equiv {B}/{2 R_c^2} \approx 
0.031 \pN $ is the elastic force scale. $L_c$ and $R_c$ are the length and 
radius of the cylindrical portion of the cell, respectively. 
If the bundle thickness $n(s)$ exhibits significant inhomogeneity, 
then the appropriate average thickness 
$n$ is the root-mean-square average thickness along the bundle length.
Other than the buckling point at $L_{fil}=L_c$, $F_B$ is 
independent of $L_{fil}$ for a given $\theta$. 
The helical pitch $p$ and pitch angle $\theta$ are related by
\begin{equation}\label{eq::pitchdefn}
  	p = \frac{2 \pi R_c}{\tan \theta}, \;\;\; 
	\cos\theta = \frac{L_c}{L_{fil}}.
\end{equation}
As the pitch $p$ vanishes,  
$\theta \rightarrow \pi/2$, and $F_B \rightarrow f_B n^2$.  The force 
$F_B$ has a maximum of $F_B^{\rm max} \equiv \frac{4}{3} f_B n^2$ at 
$\theta^* = \mbox{arctan}\, \sqrt{2} \approx 0.96 \approx 55^\circ$.

Many of our results reflect the fact that we work in a regime where 
elastic forces, $F_B$, and the polymerization force scale, $f_0 \equiv k_B T / a_0 = 
0.80 \pN$, are similar in magnitude.

\section{RESULTS}

\subsection{Stochastic steady-state}

In a typical \bsub\ cell of volume $1.9 \mum^3$, only 4 unpolymerized monomers are 
necessary to achieve the critical concentration of MreB, $c_c = 0.003 \muM$ 
--- suggesting that stochastic effects, due to the discrete molecular nature 
of the polymerizing monomers, may be significant. Similarly, the number of protofilaments 
in a typical cross-section of the MreB helix is small (less than $20$), 
and the number of protofilaments in contact with the cell wall
at the helix tips, $n_{tip}$, is even smaller.  A deterministic 
mean-field analysis of the steady-state (see Appendix~\ref{appendix::deterministic}), 
neglecting stochastic fluctuations, can be 
compared with fully stochastic simulations to explore the impact of various stochastic 
effects in this system. 

We simulated $n$ protofilaments that grew and shrank stochastically within a common 
pool of $N_0$ monomers according to the force-dependent polymerization rates, 
Eq.~\ref{eq::konf}, and force-independent 
depolymerization rates.  The forces on each protofilament tip were determined 
by the constraint force on the bundle, Eq.~\ref{eq::efm1}, divided among the leading
filaments at that tip, where $1 \leq n_{tip} \leq n$.  The filament bundle was studied once it
reached a steady-state. The length of each of the $n$ protofilaments and the 
monomer concentration continued to fluctuate within the steady-state, as did 
the number of protofilament tips, $n_{tip}$, at a given end of the filament bundle. 

\begin{figure}[h]
\includegraphics[width=8.6cm]{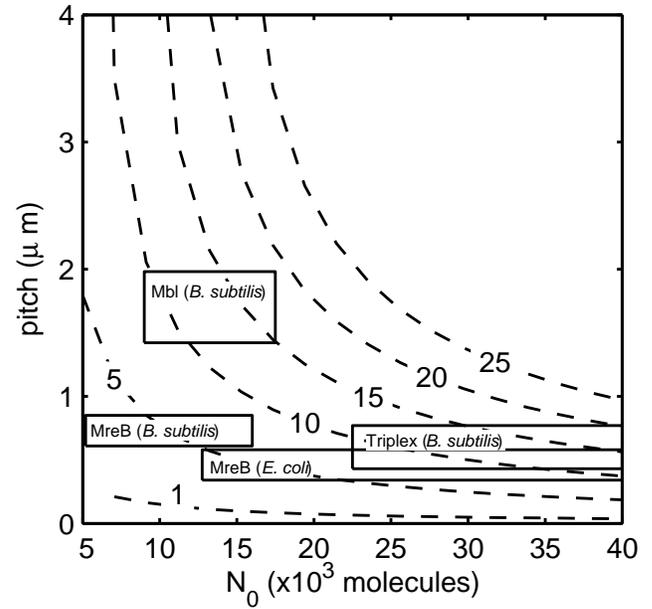} 
\caption{Steady-state pitch $p$ vs. total molecule number $N_0$ for 
various filament bulk thicknesses $n$ as predicted by stochastic 
simulation. The mean-field plot is indistinguishable from this 
due to the low cytoplasmic concentration. The rectangles represent approximate 
regions of experimental relevance from \citel{mre13, mre88, mre73}. In \bsub, if the 
three isoforms bundle together into a non-slipping triplex structure, the number of 
monomers will be the sum of each homologue.}
\label{fig::pvsn0}
\end{figure}

For a given cell geometry ($L_c$ and $R_c$) and total number of monomers ($N_0$), 
each bulk thickness $n$ yields a unique helical 
steady-state configuration with a particular pitch $p$ and average cytoplasmic 
monomer concentration $c$.  As seen in Fig.~\ref{fig::pvsn0}, as the 
abundance $N_0$ increases for a given number of protofilaments $n$, 
the pitch decreases due to longer bundles. Conversely, at a given $N_0$, 
thicker bundles (larger $n$) leads to larger pitches.  

The rectangles in Fig.~\ref{fig::pvsn0} represent independent experimental
measurements of MreB abundance and cable pitch in \bsub\ \citel{mre88,mre13} and 
\ecoli\ \citel{mre73}.  In \bsub, if the three MreB isoforms (MreB, Mbl, and MreBH) 
bundle together into a triplex structure, the total number of monomers should be the 
sum from each homologue, which we estimate is $23000 - 40000$.
We see that with current experimental pitch and abundance estimates, \ecoli\ has $n \in [3,16]$ 
protofilaments in each polarized MreB helical bundle.  Under the assumption that 
they are bundled independently of the other isoforms in \bsub\,
we estimate that Mbl has $n \in [9,21]$, while MreB has $n \in [3,7]$, with pitches from
\cite{mre13}.  Under the assumption that the isoforms are mutually non-slipping, we estimate 
$n \in [8,21]$ in the triplex with the mutual pitch from \cite{mre90}.  It is therefore possible
that the distinct pitches reported by \cite{mre13} and \cite{mre90} are the result of changes in the 
bundling properties of the variously labeled and tagged MreB analogues. 

\begin{figure}[h]
\includegraphics[width=8.6cm]{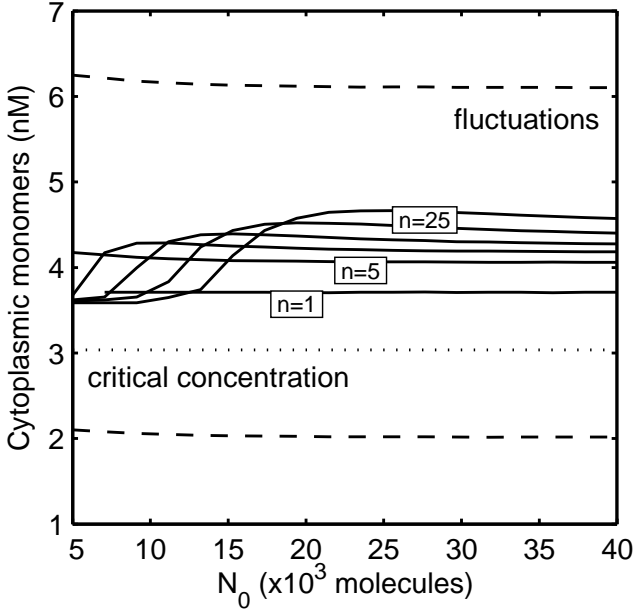}
\caption{Steady-state cytoplasmic concentration $c$ of MreB monomers  
vs. total monomer number $N_0$ from stochastic 
simulation for $n$ protofilaments, showing $n=1$, 5, 10, 15, 20, and 25.  
The dotted line represents the critical concentration. 
The dashed line illustrates the large standard deviation of stochastic fluctuations in steady-state, 
for $n=5$. For all $n$, relative fluctuations were $50\% \pm 5\%$ and absolute fluctuations
were within $0.2 nM$ of those shown for $n=5$.}
\label{fig::cssvsn0}
\end{figure}

The concentration of MreB in monomeric form 
shown in Fig.~\ref{fig::cssvsn0} is low, representing less than 
$0.1\%$ of the total cellular MreB. This is in contrast  
to the same model using eukaryotic actin kinetics, for which $1-5\%$ is 
monomeric (data not shown), and in dramatic contrast with observations 
of FtsZ polymerization, for which 30\% is associated with the Z-ring  
and 70\% is diffuse in the cytoplasm \invivo\ \citel{mre89}.  Notwithstanding, the average monomeric 
concentration of MreB is significantly above the critical concentration due to a 
reduced polymerization rate which arises from the constraining force at the bundle tips.
The shape of the curves in Fig.~\ref{fig::cssvsn0} follows from the force vs. pitch relationship in
Eq.~\ref{eq::efm1}, which grows quickly, reaches a maximum then decreases again as the pitch
angle approaches $\pi/2$. As we shall see, force and cytoplasmic concentration increase together, 
while, as we have seen, pitch and $N_0$ are inversely related. As a result, for each $n$ 
the cytoplasmic concentration exhibits a similar maximum vs. $N_0$ as force does vs. pitch. 
This maximum, corresponding to $\theta^\ast$, is at larger $N_0$ for larger $n$.
We also see in Fig.~\ref{fig::cssvsn0} 
that the fluctuations in the monomeric concentration are very large and approximately 
independent of $n$ and $N_0$.  The cell is often instantaneously 
below the critical concentration despite the upward bias due to the constraint forces. 

As seen in Fig.~\ref{fig::fvscss}a, the maximal force $F$ sustainable by the 
bundle in steady state increases with the cytoplasmic concentration $c$.  
For this part of the figure $c$  was held fixed, to facilitate comparison with 
both the mean-field calculations in Appendix~B and the stochastic $n=2$ calculations 
in Appendix~C.  The increase
of $F$ with $c$ comes from Eq.~\ref{eq::konf}, the force-induced
reduction of the polymerization-rate, and reflects the excess of the monomer
fraction over the critical concentration line seen in Fig.~\ref{fig::cssvsn0}.
$F$ increases with the number of protofilaments, $n$, at a fixed
cytoplasmic concentration, since it is distributed over the leading 
$n_{tip}$ protofilaments. 
The excess of $F$ over the {\em maximal} mean-field prediction (dashed lines), where $n_{tip}=n$,
is recovered analytically for $n=2$ (solid line) in Appendix~\ref{appendix::stochastictip}. 
The excess arises because any of the $n_{tip}$ protofilaments may grow and extend
the bundle length, after which the remaining near-tip protofilaments will quickly catch up 
without force retardation.  The result is enhanced growth at a given $F$, or, equivalently, 
a larger $F$ at which steady-state is reached.  The stochastically-enhanced force is more significant 
as $n$ increases, and also increases with $c$ due to the increased polymerization rate. 

In Fig.~\ref{fig::fvscss}b, we examine the additional effects on $F$ due to the large
fluctuations in $c$ that were apparent in Fig.~\ref{fig::cssvsn0}.
Here we plot $F_B$ vs. the average $c$ for fully fluctuating filament
bundles. Various $c$ were explored by varying $N_0$.
We include the mean-field results (dashed lines) for reference.  Two differences with Fig.~\ref{fig::fvscss}a 
are apparent. The first is that the fully fluctuating results have a maximal force 
sustainable by the elastic bundle,
$F_B^{\rm max}=4/3 f_B n^2$. With $c$ fixed we could use an arbitrarily stiff
bundle to explore a wider range of forces, but here we need to choose a 
particular stiffness (by equating $F=F_B$) to accurately couple fluctuations in $c$ with those of $F$.  
The second difference is that $F$ for the fully-fluctuating system is {\em below} the mean-field results for 
$n \leq 5$, and is lower than the results in Fig.~\ref{fig::fvscss}a for a fluctuating $n_{tip}$. 
This systematic decrease can be seen as arising from averaging the concave-down curves from Fig.~\ref{fig::fvscss}
over the very large $c$ fluctuations, which leads to a stronger decrease for smaller average $c$. 

There are significant effects on the forces sustained by the filament bundle both due to 
fluctuations in the small number of protofilaments supporting the force at the bundle tip, $n_{tip}$, and 
due to the large fluctuations in the monomer concentration, $c$. Because of the curvature of the 
$F$ vs $c$ curve, these fluctuation effects modify the mean-field steady-state bundle force in 
opposite directions.  While for small numbers of protofilaments ($n \lesssim 5$) the overall effect 
is to reduce $F$ for a given $c$, the net effect for the physiological range of bundle thicknesses 
(Fig.~\ref{fig::pvsn0}) and cytoplasmic concentrations (Fig.~\ref{fig::cssvsn0}) is an increased force. 

\subsection{Cell growth}
\label{growth}

The MreB helix grows as the cell grows, doubling its lateral length before dividing. 
Our quasi-static approach can accommodate cell 
growth. We assume that the total number of monomers $N_0$ is 
proportional to the cell length and that the number of protofilaments $n$ is 
length-independent. We 
simulated cell growth in a regime towards the low-$N_0$ end of Fig.~\ref{fig::pvsn0},
with $N_0 = 3\times 10^3 L_c$ so that an average $3\mum$ 
cell contains $9000$ MreB monomers, and towards the high-$N_0$ end, with 
$N_0 = 10^4 L_c$.  Fig.~\ref{fig::growth} shows how the steady-state helical pitch 
varies as the cell length, $L_c$, ranges between $2 - 4 \mum$. 

For most bundle thicknesses, $n$, the pitch is nearly constant as the cell 
elongates. However, for thicker bundles in the low-$N_0$ regime, the 
helical pitch increases significantly as the cell doubles in size.  This 
increase is due to fluctuations at small cell lengths, $L_c$.  
While it appears strange that at {\em large} $n$, the stochastic
effects are larger, the pitch is determined by the filament length $L_{fil}$, which 
only depends on the maximum protofilament length. While the mean 
protofilament length fluctuates less with increasing $n$, the maximum 
protofilament length is an extremal property of the bundle --- and 
increases with increasing $n$.  At small $L_c$, the relative fluctuations in the cytoplasmic 
fraction also increases.  For the larger $N_0$ regime of Fig.~\ref{fig::growth}b,
the pitches are much smaller, $L_{fil}$ larger, and relative length fluctuations correspondingly smaller.

In the physiological regimes shown by boxes in Fig.~\ref{fig::pvsn0},
pitch should not change significantly during cell growth if the overall concentration of 
MreB monomers remains constant. 
Mbl in \bsub\ may have longer pitches in longer cells \citel{mre25, mre29}, 
though this is not a strong effect \citel{mre13}.  
Experimental observations of considerable variability of 
the number of helical turns per cell within cells of the same
strain, size and growth conditions \citel{mre25} may imply corresponding variability
of MreB expression or of the cross-sectional number of protofilaments $n$ --- which complicates
analysis. 

\subsection{Macromolecule trafficking}

A vital role of MreB is the polar localization of proteins such as 
Tar in \ecoli\ \citel{mre37}, the cell polarity markers DivJ and PleC in 
\caul\ \citel{mre34}, and the origin-proximal regions of the 
newly-replicated chromosome in \ecoli, \bsub\ and \caul\ \citel{mre50}. 
One possibility is that these passengers associate with yet-to-be-discovered 
motor proteins that use MreB as a track to the poles \citel{mre37}. A 
second possibility is that these proteins simply bind to the helix and 
advect with the continuous treadmilling, eventually ending up at one of the 
polar tips.  A third possibility is that they associate with leading tips of MreB 
protofilaments \citel{mre33}, perhaps via intermediary proteins 
analogous to formin for actin filaments.  Here we quantitatively analyze these 
possibilities, which are depicted in Fig.~\ref{fig::translocationcartoon},
together with associated translocation speeds with respect to the fixed bacterial 
axis.

For any transport mechanism, a characteristic speed along the filament
bundle, $v_{trans}$, translates into a speed 
$v_{z} = \cos\theta v_{trans}$ relative to the cell's axis. 
Assuming the protein initially binds at a uniformly 
random location along the bundle, the average time to reach a pole is then
\begin{equation} \label{eq::ttrans}
 	\langle t_{trans} \rangle \approx \frac{L_{c}}{2 v_{z}} 
		= \frac{L_c}{2 \cos\theta v_{trans}}. 
\end{equation}
This time can be compared to the cell division time to see whether it provides
a plausible mechanism for polar localization. 

The myosins that transport organelles along actin tracks in 
eukaryotes travel at speed $v_{mot} \approx 200-400 \nm/\s$ \citel{act25}. 
Attached to putative myosin homologues, macromolecules could be 
translocated to the poles in $\sim 20$ seconds. This is well within cell 
division times, so this would be a viable polarization mechanism. However, 
no cytoplasmic motor homologue has been identified in prokaryotic cells. 
Furthermore, almost all myosins travel towards the pointed (``$+$'') tip along
actin filaments \citel{act28}, so a single polarized MreB bundle 
would probably only support motor-driven localization to one pole while
unpolarized bundles would not support selective targeting to one pole
and not the other.  

Static association of proteins to the side of MreB bundles would, through
treadmilling, lead to a translocation speed equal to the rate 
of advection times the monomer spacing, $\lambda_{tread} a_0$,
in the direction of the pointed end of the bundle. In steady state, the
advection rate $\lambda_{tread} = 1.2 \s^{-1}$ 
is independent of the buckling force, the bundle thickness,
or the cytoplasmic concentration  and leads to 
$v_{tread}= a_0 \lambda_{tread}= 5.92 nm/s$, shown by
the dashed line in Fig.~\ref{fig::translocation}c. Applying Eq.~\ref{eq::ttrans} to a 
typical configuration with $p = 1 \mum$ and $L_c = 3 \mum$, this yields 
$\langle t_{trans} \rangle \approx 12$ minutes, which is plausible 
compared to cell-division timescales and comparable to {\em oriC} translocation times
\cite{mre91}. 

Proteins that could bind either directly, or via putative tip-binding proteins
(analogous to formin in eukaryotes \citel{mre33,act27})
to the barbed end of free protofilaments could be translocated in the {\em opposite}
direction to the treadmilling advection.  In an unbuckled ($F_B=0$) filament, lateral 
protofilaments treadmill at the same rate of the net backwards advection 
of the bundle, accomplishing no net movement relative to the cell's axis. 
In a buckled filament, however, the monomer concentration is considerably above 
the critical concentration of a free protofilament, as illustrated in 
Fig.~\ref{fig::cssvsn0}. As a result, unconstrained laterally associated barbed 
ends grow faster than the bulk of the bundle, with 
\begin{eqnarray}
 	v_b &=& a_0 \left[ (\kbon c - \kboff) - \lambda_{tread} \right] \notag\\
 	&=& a_0 \kbon (c - c_c).
	\label{eq::barbedassoc}
\end{eqnarray}
For each bundle thickness and pitch simulated, the translocation times are 
shown in Fig.~\ref{fig::translocation}a and speeds in Fig.~\ref{fig::translocation}c. 
Typical bundles with $20000-40000$ molecules of MreB and $8-15$ protofilaments 
thick would transport passengers in $4-6$ minutes, considerably faster than 
laterally-associated proteins and in the opposite direction.  

Proteins associated with the slow-growing pointed end have a 
net velocity given by 
\begin{eqnarray}
 	v_p &=& a_0 \left[ (\kpon c - \kpoff) + \lambda_{tread} \right] \notag\\
 	&=& a_0 \kpon (c - c_c).
	\label{eq::pointedassoc}
\end{eqnarray}
Note that free pointed ends are {\em disassembling} on average, though not as rapidly
as the treadmilling, so that $0< v_p < v_{tread}$.  
These translocation times are shown in Fig.~\ref{fig::translocation}b and speeds
in Fig.~\ref{fig::translocation}d. According to 
Eqs.~\ref{eq::barbedassoc} and~\ref{eq::pointedassoc}, they are 
slower by a factor of $\kbon / \kpon \approx 10$ compared to free barbed ends, 
taking several hours.  These times are probably too slow to be biologically relevant.

These protofilament-associated translocation modes offer a non-motor based mechanism 
for specific targeting of proteins to {\em either} pole in cells with a single 
polarized bundle of MreB.  The 
cell could specify the specific pole destination of a particular protein by 
specifying which part of an MreB protofilament it binds to: 
barbed-associated proteins would end up at the pole at the barbed-tip of the MreB
bundle with a speed of $v_b$, while laterally-associated proteins would end up at 
the pointed-tip pole with speed $v_{tread}=a_0 \lambda_{tread}$.  Of course, these
translocation mechanisms may also supplement a (hitherto undiscovered) motor-based
mechanism to provide targeting to either pole with polarized MreB bundles. We do not
see any way of specific targeting of proteins to a given pole if the MreB bundle is
not polarized or if there are anti-parallel bundles, either with or without motor 
proteins. 

If protofilaments dissociate at a significant rate from the main bundles then 
these translocation times represent lower bounds. Additionally, any putative 
MreB-binding proteins could strongly affect the 
polymerization kinetics. For example, ADF/cofilin in eukaryotes increase $\kpoff$ for
actin by $\sim 20$ times. A bacterial homologue of such a protein would decrease the 
delivery time of pointed-associated proteins by $\sim 20$. Similarly, in the presence of 
profilin, formin increases the barbed growth rate of actin by $10$- to $15$-fold 
\citel{mre27}, and such a modification 
would decrease the delivery time of barbed-associated proteins to within a minute. 

Proteins associated with pointed or barbed ends of 
protofilaments could be translocated towards those poles at $v_p$ and $v_b$, respectively. However
the proteins could also be directly recruited to distinct poles of the cell, due to the free
ends of the MreB bundles. The relative
magnitude of translocation vs. direct recruitment is dependent on the number of barbed or pointed 
protofilament ends. Tip-associated translocation requires a significant number of
laterally associated protofilaments tips to maximize the translocation flux, 
while tip-directed polar recruitment
requires unbroken protofilaments to minimize non-polar binding sites.  Observations
in \caul\ \citel{mre71} indicate that protofilaments are short, supporting
tip-associated translocation as a viable mechanism {\em in vivo}. 

\subsection{Recovery dynamics of helices}

In experiments on \bsub\ involving fluorescence recovery after photobleaching 
(FRAP) of fluorescence-tagged Mbl, four helical turns on one side of the 
cell's longitudinal axis were bleached while the other half continued to 
fluoresce. It took approximately $8$ minutes for the bleached halves to 
recover fluorescence at the same level as the unbleached halves \citel{mre29}. We obtain 
an upper bound for this time by calculating the average time unbleached parts of 
the MreB protofilaments take to treadmill into the bleached regions; one half-turn 
in the polarized model and one quarter-turn in the non-polarized models. 
Thus, in silico,
\begin{eqnarray}
 	t_{FRAP} \approx \frac{\pi R_c}{a_0 \sin\theta\lambda_{tread}} 
		= \frac{\pi \sqrt{p^2 + R_c^2}}{a_0 \lambda_{tread}}
\end{eqnarray}
for polarized array structures and half of that for non-polarized array 
structures. For $p \approx 0.5 \mum$ this yields between 3 and 6 minutes. 
The agreement of this timescale with experiment suggest that monomer renewal 
by exchange with the cytoplasm may not dominate the FRAP recovery time. 

Several experiments have applied the MreB-specific small molecule A22 to 
quickly and reversibly break down the MreB cytoskeleton by blocking 
polymerization \citel{mre26,mre50}. Cells remain viable after recovery from A22-induced 
disruption of MreB and reform their helical patterns 
in less than 1 minute for \caul\ \citel{mre26}.  
In the context of our model, recovery 
from A22-treatment corresponds to re-establishment of the 
steady-state MreB helix from a pool of cytoplasmic monomers. We simulated 
our stochastic model from a nucleus of $n$ protofilaments, each of 
length 3 (as suggested for actin \citel{act21}), under the assumption that 
the nucleation time is short \citel{mre38}. Thicker 
bundles (larger $n$) reach their final steady-state length much faster than thin 
bundles, due to the presence of more free filament ends. The 
equilibration times vary between $1-5$ seconds. Fewer bundled 
protofilaments took longer to reach a steady state, the longest being 
$n=1$, with $N_0=10000$, in which the final length is reached in $5$ 
seconds.  These are consistent with A22 recovery timescales. 

We can also address the timescale of breakdown. Assuming that A22 simply blocks polymerization, but 
does not change depolymerization dynamics, then it will take 
$t_1=L_{fil}/(k^p_{off}+k^b_{off}) \approx 13$ minutes for each protofilament to disassemble with 
no internal free ends. If each protofilament has length $L_{proto}$ with $m=L_{fil}/L_{proto}$ 
free pointed (or barbed) ends, then we would expect
the disassembly to be correspondingly faster by $t_m = t_1/m$. Experiments constrain the actual 
disassembly time in \caul\ \citel{mre26} to be $\leq 1$ min, implying $m \geq 13$ and 
$L_{proto} \approx 300 \nm$.  Analysis of single-molecule experiments in \caul\ 
estimates $L_{proto} \approx 392 \pm 23 \nm$ \citel{mre71} by assuming that protofilaments 
treadmill in place.  If protofilaments also advect along the bulk MreB cable, as we assume in our model, 
this experimental estimate represents an upper bound and is consistent with our result.

These estimates of assembly and disassembly timescales of the MreB helix can also be applied to the 
reported midcell condensation of MreB in  \ecoli\ \citel{mre45} and \caul\ \citel{mre23}, 
and of MreBH in \bsub\ \citel{mre89}.  Two possible mechanisms are
an elastic compression of the intact MreB helix to midcell driven by some (posited) motor protein, 
or disassociation of the MreB helix and (transient) association with some midcell binding 
partners.  These mechanisms may occur in tandem. 
The timescales of condensation and recovery are $\approx 30$ minutes in \caul\ which is 
considerably longer than A22-induced breakdown. The maximum 
filament-end force needed to compress a typical MreB helix in our model from $L_c$ to a midcell spiral is 
$F_B^{\rm max} \approx 10 pN$, and is comparable to the force generated by the RNA polymerase \citel{mre50} 
and the anchoring forces of integral membrane proteins \citel{act29}. Thus, both motor driven 
compression and depolymerization are plausible mechanisms for the observed midcell condensation. 

\section{DISCUSSION}

Our model explains how polymerization forces in a bundle of MreB 
protofilaments can maintain a helical configuration in 
mechanical equilibrium. Helices are observed increasingly often within 
bacteria \citel{mre92} and our model suggests an elastic mechanism for maintaining 
such helicity when filament bundles extend the length of the cell. 
For MreB, our model predicts that the observed pitches and protein 
abundance are consistent with bundles of $10-20$ protofilaments thick.

Helical pitch is directly observable through fluorescence microscopy, and the 
total abundance of MreB can be controlled by an inducible promoter \citel{mre29}.  
Increasing the MreB expression level should systematically
decrease the pitch according to the curves in Fig.~\ref{fig::pvsn0}.
We have assumed a cytoplasmic pool of monomers, while 
there is evidence of an oligomeric pool of Mbl in \bsub\ \citel{mre29}. Since
only monomers will contribute to force-generation through polymerization, an oligomeric pool that
was not part of the MreB bundle would simply
lead to a shift along the $N_0$ axis in Fig.~\ref{fig::pvsn0} and \ref{fig::cssvsn0}.

Filament treadmilling is ongoing in the steady-state.
A significant result is that parts of the filament may travel 
faster than the bulk treadmilling rate, even in the opposite direction, 
while other parts may travel much slower. Since the cytoplasmic concentration 
is held above the critical concentration of free protofilaments by the constraining forces 
at the filament tips, the barbed end of protofilaments that associate 
laterally with the MreB bundles can travel at speeds greater than the bulk 
advection speed. These speeds are fast enough to be biologically relevant 
in the transport of polar-targeted proteins and the origin-proximal part 
of the chromosome. These transport modes may explain how MreB cables 
transport passengers without motor proteins, and may also reconcile 
experimental observations of very different speeds of MreB protofilaments: 
fast speeds of $0.07 \mum/\s$ \citel{mre25} may correspond to 
the lateral protofilaments described above, while slow speeds 
comparable to actin treadmilling speeds \citel{mre71} may correspond to 
protofilaments involved in the bulk of the cables.  It also suggests that observations of 
bidirectional transport along MreB cables \cite{mre71} does not necessarily rule
out a single polarized MreB bundle.

It has been suggested that cell polarity arises from the inherent polarity 
of MreB polymers \citel{mre33, mre34}. The different modes of transport in 
our model provide a mechanism for translating the polarity of a cable
to the entire cell. We propose that proteins destined for one pole 
associate laterally with the cable, and are naturally translocated 
towards the ``pointed" pole. Proteins destined for the other pole bind to 
the barbed ends of free protofilaments and are naturally translocated 
towards the ``barbed" pole. We have demonstrated that such translocation 
happens within a physiologically relevant timescale. Such selective 
targeting requires polarized MreB filaments. Proper 
polarization of the MreB cables themselves may require an upstream 
nucleator, for which TipN in \caul\ is a candidate \citel{mre48}.  While antiparallel
MreB cables can support transport to both poles, we do not see how it can support
targeted transport to specific poles. 

Evidence for polarity of MreB in \caul\ is mixed \citel{mre34,mre71}, however 
both PleC and DivJ appear to have targeted localization to specific poles during 
the cell cycle \citel{mre50}. This is consistent with a single polarized MreB cable in \caul.
We are not aware of evidence about MreB cable polarization in \ecoli. While
FRAP recovery of Mbl cables in \bsub\ is symmetric \citel{mre29}, this could 
be due to lateral exchange and/or the motion of laterally associated protofilaments. 

Any transport along treadmilling MreB bundles would require either direct association
of targeted proteins with MreB, or association through an intermediary. 
Interactions of only a few cytoplasmic proteins with MreB have been identified. 
These include SetB, implicated in chromosome segregation \citel{mre53}, RNA polymerase 
\citel{mre50}, and the chaperonin GroEL \citel{mre50}.  Of these, GroEL is known
to interact with a large number of cytoplasmic proteins 
\citel{mre93}, while RNA polymerase is needed for their 
transcription. GroEL is localized to polar and septal 
regions in \ecoli\ \citel{mre96} while RNAP is associated with nucleoids in 
punctate patterns similar to those seen for helically distributed proteins \cite{mre94, mre95}.
It would be interesting to see if any of these MreB associated cytoplasmic proteins 
exhibit pole directed transport within bacterial cells. 

Transport speeds may be decreased if short protofilaments necessitate 
multiple binding and release events to translocate from 
one pole to the other.  Tip-directed transport also allows for direct recruitment to free
protofilament tips, such as at the bundle ends. 
The details of bundle ultrastructure and of protein translocation
will be required to sort out these competing effects. 

We have shown that many of the dynamical phenomenon associated with MreB, including polar 
protein localization, MreB helix dissolution and reformation, and FRAP studies, could result
from polymerization and treadmilling of bundled MreB protofilaments without motor proteins.  
In order to make precise quantitative predictions, direct measurements of 
the kinetic rate constants for MreB polymerization would be invaluable. 
The \therm-scaled MreB parameters that we used are consistent with experiment and require the least 
amount of {\em ad hoc} parameter adjustment.  
Pitch-abundance and concentration-abundance relationships 
did not change in character between eukaryotic actin and our \therm-scaled MreB kinetics. 
Conversely, translocation times and timescales for dissolution and reformation of MreB helices 
are strongly parameter-dependent.  However, our qualitative result of bidirectional transport 
is general since, from Eq.~\ref{eq::barbedassoc}, $v_b$ is always in the opposite direction 
from the bulk treadmilling. 

There are two significant sources of stochastic effects in our model.  
Stochastic fluctuations due to the number of force-bearing protofilaments, 
$n_{tip}$, significantly increases the net force applied to the bundle (Fig.~4A).   
There are also large fluctuations in the cytoplasmic monomer concentration in the steady-state, 
$c$ (Fig.~3), and these systematically {\em decrease} the net forces (Fig.~4B). We have shown that
the stochastic effects of $c$, due to the small cell volume combined with the low critical 
concentration, and the stochastic effects of $n_{tip}$, due to the relatively small number of 
protofilaments, are both significant.

We have assumed that MreB polymers are naturally straight, without intrinsic curvature or twist. 
There are suggestions that MreB helices are always right-handed \citel{mre13}. Handedness could
result from nucleation conditions or constraints at the helix ends, which would not affect our results. 
Handedness could also arise from a non-zero intrinsic twist $\tau_o$ in Eq.~\ref{eq::hamiltonian}. 
For small $|\tau_o| \ll R_c^{-1}$, the elastic force in Eq.~\ref{eq::efm1} would remain unaffected. 
It has been hypothesized that MreB exhibits an anisotropic membrane affinity requiring one face of 
the monomers to always orient towards the membrane together with significant intrinsic curvatures
\citel{mre104}.  If so, our polymerization dynamics could still be applied though with a more 
complex energy functional \cite{mre104}. However, 
we would expect qualitatively similar results if helical extension remains force-limited.

We have demonstrated that many of the dynamical phenomena associated with MreB including polar 
protein localization, MreB helix dissolution and reformation, and FRAP studies, can be 
addressed by polymerization and treadmilling of bundled MreB protofilaments without the need to 
invoke motor proteins.  The study of the MreB helix ultrastructure, as well as
the detailed study of the translocation of MreB-associated pole-directed proteins in the various 
bacterial species with various MreB analogues, will shed further light on the viability of our
results {\em in vivo}.  

\acknowledgments

We thank NSERC for financial support, and Rut Carballido-Lopez for discussions.

\appendix
\section{Global elasticity model}\label{appendix::globalelastic}

Following Antman \citel{ela01}, an elastic filament is described by the position 
of its centerline $\vec{r}(s)$ and an orthonormal basis of directors 
$\{\vec{d}_1(s), \vec{d}_2(s), \vec{d}_3(s)\}$ specifying the orientation 
of its cross-section. This approach has been used extensively for DNA 
\citel{ela14}, but is seldom used for eukaryotic actin (though see 
\citel{ela15}) since  actin does not systematically form structures {\em in vivo} that 
are much smaller than its persistence length. In contrast, MreB forms a 
helix with a radius of $\approx 400 \nm$, which is smaller than the 
persistence length of a single protofilament ($\xi_p = B/k_B T \approx 15 
\mum$) and much smaller than $\xi_p$ for a bundle of several 
protofilaments. 

The stretching modulus of a filament bundle of actin is $n E \approx 40 
n \pN$ where $n$ is the number of protofilaments in a typical 
cross-section of the bundle. For a typical filament with $n=15$, the 
energy scale to stretch the filament by monomer addition is 
$\Delta U_{stretch} = E\, n a_0 \approx 
2\times 10^3 \pN \nm$, whereas a comparable energy scale for 
bending the helix is $\Delta U_{bend} = {B a_0}/{4 R_c^2} 
n^2\approx 70 \pN \nm$. We therefore assume that the MreB filament bundle is 
inextensible and parametrize the filament by its 
arclength $s \in [0,L_{fil}]$. The angular strains in the 
filament are then simply $\ds\vec{d}_i = \vec{u} \times \vec{d}_i$, $\vec{u} 
= \kappa_1 \vec{d}_1 + \kappa_2 \vec{d}_2 + \tau \vec{d}_3$ where 
$\kappa^2 \equiv \kappa_1^2 + \kappa_2^2$ is the curvature and $\tau$ is 
the local twist.  We set $\vec{d}_3 \equiv \ds \vec{r}$, following the 
standard shear-free assumption of biopolymers. 

The centerline of a helix with pitch angle $\theta$ (measured from the 
cell's axis, $\hat{z}$) and radius $R_c$ is
\begin{multline} \label{eq::helix}
 	\vec{r}(s) = R_c \cos\left( \frac{\sin\theta}{R_c} s \right) \hat{x}
 	+ R_c \sin\left( \frac{\sin\theta}{R_c} s \right)  \hat{y}\\
 	+ \cos\theta s  \;\hat{z}.
\end{multline}
Using the tangent, normal and binormal unit vectors, $\{\vec{t}(s), 
\vec{n}(s), \vec{b}(s)\}$, and the Frenet-Serret Theorem 
\citel{ela01} we obtain the curvature of the helix 
\begin{equation}
 	\kappa = \frac{\mbox{sin}^2{\theta}}{R_c},
\end{equation}
and the torsion of the filament's centerline, 
\begin{equation}
 	\tau_c = \frac{1}{R_c}\cos{\theta}\sin{\theta}.
\end{equation}
The directors $\{\vec{d}_1(s), \vec{d}_2(s)\}$ are, in general, a rotation 
of $\{\vec{n}(s), \vec{b}(s)\}$ through an angle $\phi(s)$.  The 
difference between the physically relevant total twist $\tau$ and the 
centerline torsion $\tau_c$ is $\tau_L \equiv \tau-\tau_c= \ds \phi$ \citel{ela13}. 

Using the linearly elastic Hamiltonian from Eq.~\ref{eq::hamiltonian} and 
$L_c = L_{fil} \, \cos\theta$ and assuming that 
$\tau_L$ is free to rotate to eliminate $\tau(s)$, we find
\begin{eqnarray}
 	\frac{\partial \mathcal{H}}{\partial L_c} &=& 
		- \frac{2 B}{R_c^2} \;\mbox{sin}^2\theta \cos\theta 
			\langle n^2 \rangle \label{eq::varlc},\\
 	\frac{\partial \mathcal{H}}{\partial R_c} &=& 
		- \frac{B L_{fil}}{R_c^3} \;\mbox{sin}^4 \theta 
			\langle n^2 \rangle \label{eq::varrc},\\
 	\frac{\partial \mathcal{H}}{\partial L_{fil}} &=& 
		- \frac{B}{2 R_c^2}\; \mbox{sin}^2 \theta\; 
		(1+ 3\, \mbox{cos}^2\theta ) 
			\langle n^2 \rangle \label{eq::varlfil}.
\end{eqnarray}

Regardless of any mechanisms holding the MreB bundle in a helical configuration, 
one additional monomer must provide an energy of 
$ \left( - a_0 {\partial \mathcal{H}}/{\partial 
L_{fil}} \right)$ to polymerize itself to the tip of the longest protofilament(s). 
This provides an estimate for the force acting upon the tip of the filament bundle,
\begin{equation}
  	F_B = \begin{cases}
 	f_B \;\mbox{sin}^2 \theta\; (1+ 3\,\mbox{cos}^2\theta )  
		\langle n^2 \rangle & L_{fil} > L_c \\
 	0 & L_{fil}< L_c,
  \end{cases}
  \end{equation}
where $f_B \equiv {B}/{2 R_c^2} \approx 0.031 pN$. This implies that $F_B \approx 10 pN$
for $n \approx 10$.  This is comparable to pull-out forces 
of integral membrane proteins \citel{act29}.

Eq.~\ref{eq::varlc} 
provides an estimate for the force generated by the MreB cytoskeleton 
against the cell's end caps. An estimate for the difference in 
longitudinal spring constant of the cell with and without a properly 
formed MreB helix is,
\begin{equation}
  \Delta k_{cell} \approx  \frac{\partial^2 \mathcal{H}}{\partial L_c^2} 
	= \frac{2 B}{L_c R_c^2} \cos\theta (3\cos\theta - 1) n^2.
\end{equation}
For $p = 1 \mum$ and $n=15$, this yields $4 \times 10^{-4} \pN / \nm$. 
For \ecoli, the spring constant of the entire cell is $\approx 10^3 
\pN/\nm$ \citel{mre49} and is much larger for \bsub, thus the spring 
constant differential provided by the MreB helix is insignificant.
Similarly, Eq.~\ref{eq::varrc} provides an estimate for the radial line 
pressure the helix exerts on the lateral walls:
\begin{equation}
  \frac{F_{\hat r}}{L_{fil}} \approx \frac{B}{R_c^3}  
	\;\mbox{sin}^4 \theta n^2. 
  \label{eq::afm}
\end{equation}
For $p = 1 \mum$ and $n=15$, this yields $0.03 \pN / \nm$. 
Eq.~\ref{eq::afm} represents a force contrast between pushing the cell 
wall directly above the helix and elsewhere. A 
bundle with $n=15$ has approximate thickness $a_0 \sqrt{n} \approx 20 \nm$ 
and the local excess pressure over the cables is $\approx 5 \times 10^{-3} 
\pN/\nm^2$. In comparison, turgor pressure inside \ecoli\ has been 
measured at $\approx 0.1 \pN /\nm^2$ \citel{mre49}, indicating that the 
rigidity of the MreB bundles does not directly provide significant structural 
support for the cell well.

We may also use the elastic model to check the self-consistency of our 
quasi-static helical configuration, Eq.~\ref{eq::helix}.  
At physiological temperatures, all cellular components undergo thermal 
fluctuations of the order $k_B T$. However, a typical helical bundle with 
$n = 5$ and $p = 1 \mum$ has an elastic energy of $U = 3.5 \times 10^5 \pN 
\nm \gg k_B T = 4.1 \pN\nm$. We therefore assume that the helical pitch does 
not significantly fluctuate due to thermal effects.  An 
estimate for the timescale for elastic reorganization of the MreB helix 
can also be obtained from the relaxation time of the lowest hydrodynamic mode in 
the filament \citel{act30}, 
\begin{equation}
  t_{elastic} \sim \frac{2^6}{\left(3 \pi \right)^4 \mbox{ln}2} 
	\frac{\eta L_{fil}^4}{B} \approx 0.02 \s,
\end{equation}
where $\eta$ is the cytoplasmic viscosity. In comparison, the characteristic 
time for polymer elongation is given by the steady-state treadmilling rate,
\begin{equation}
  t_{poly} \sim \frac{1}{\lambda_{tread}} = 
		\frac{\kpon + \kbon}{\kpoff\kbon - \kboff\kpon} \approx 1.7 \s.
\end{equation}
Since $t_{elastic} \ll t_{poly}$ we can assume that elastic relaxation is fast
compared to the polymerization dynamics of interest. 

\section[deterministic]{Mean-field steady-state}
\label{appendix::deterministic}

By neglecting stochastic fluctuations, we can analytically relate pitch, 
total monomer number and cytoplasmic concentration. The two essential ingredients are 
stationary filament lengths due to treadmilling in the presence of elastic constraint
forces, and conservation of the monomer number.  
If $n_{tip} \in [1, n]$ protofilaments reach each tip, then from 
Eqs.~\ref{eq::konf}, \ref{eq::ccdef}, and \ref{eq::efm1},
\begin{equation} \label{eq::fstall}
 	F_B = f_0 \; \mbox{ln} \left( \frac{c}{c_c}\right) n_{tip} = 
		f_B\; \mbox{sin}^2\theta 
		\;(1+3 \mbox{cos}^2 \theta) \langle n^2 \rangle.
\end{equation}
We also have conservation of total number of monomers,
\begin{equation}
 	N_0 = N_A c + \langle n \rangle \frac{L_c}{a_0} \frac{1}{cos \theta},
	\label{eq::conserve}
\end{equation} 
where $N_0$ is the total number of monomers, $N_A$ is the number of molecules
per $\muM$ for a given cell volume, and $\langle n \rangle$ is averaged along the length of the 
helix. Together these equations give a relationship
between concentration and pitch angle,
\begin{multline}
 	\mbox{ln} \left(\frac{c}{c_c}\right) n_{tip} = 
		\frac{f_B}{f_0} \mbox{sin}^2\theta \;
		(1+3 \mbox{cos}^2 \theta) \\
 	\times \left( \left( N_0 - N_A c\right)^2 \frac{a_0}{L_c}^2 
		\;\mbox{cos}^2 \theta\right),
	\label{eq::MFconc-pitch}
\end{multline}
and an equivalent relationship between bulk thickness and pitch angle,
\begin{multline}
 	\mbox{ln}\left( \frac{N_0}{N_A c_c} - 
		\langle n \rangle \frac{L_c}{a_0 N_A c_c}\frac{1}{\mbox{cos}
			\;\theta}\right) n_{tip} \\
 	= \frac{f_B}{f_0} \mbox{sin}^2\theta 
			\;(1+3 \mbox{cos}^2 \theta) \langle n^2 \rangle.
	\label{eq::MFthick-pitch}
\end{multline}

The mean-field force-concentration relation obtained
from Eq.~\ref{eq::fstall} and Eq.~\ref{eq::MFthick-pitch} is plotted in 
Fig.~\ref{fig::fvscss} for $n_{tip}=n$, which maximizes $F_B$ for the bundle. 

\section{Stochastic bundle tip, {\MakeLowercase{$n_{bulk} = 2$}}}
\label{appendix::stochastictip}

If a force $F$ is applied to both ends of a protofilament, both $\kbon$ 
and $\kpon$ will be reduced by a factor of $\epsilon \equiv e^{-F/f_0}$ \citel{act01, 
act20}. Though the treadmilling rate is unaffected by the forcing, the treadmilling 
concentration becomes
\begin{eqnarray} \label{eq::ccforced}
 	c = \left(\frac{\kboff+\kpoff}{\kbon+\kpon}\right) e^{F/f_0}.
\end{eqnarray}

In the absence of forces, every protofilament of a bundle treadmills with 
equivalent rates and critical concentrations.
However, force generation by a bundle in which the load is applied only on
the most advanced $n_{tip}$ protofilaments behaves differently. 
Lagging filament tips grow without force retardation and with a reduced treadmilling
concentration, thereby catching up to the tip more often than otherwise --- increasing
$n_{tip}$ and reducing the load per loaded filament. This leads to a larger total
load for the bundle for the same monomer concentration $c$.  A similar effective 
increase in force generation due to stochastic fluctuations has been noted in passing 
\citel{act03}, although not analyzed in detail. Here we explicitly work out the
details for $n=2$.  For $n > 2$, the effects of force generation by a stochastic $n_{tip}$ are 
expected to be even more significant, as is seen in Fig.~4. 

We parametrize the system of two filaments by $i \in \{0,\,1,\,...\}$, the 
number of monomer spacings between the two tips. For any $i$, two 
competing Poisson events could increase the spacing: addition at the 
leading tip and dissociation at the lagging tip. Reduction of $i$ occurs 
by the complementary two events. The master equations lead to the following equations for stationary 
probabilities $p_i$:
\begin{multline} \notag
 	i \geq 2 :\: p_{i-1} ( \koff + \kon c \epsilon) 
		+ p_{i+1} ( \kon c + \koff ) \\
 	= p_i (\koff + \kon c \epsilon + \kon c + \koff ),
 	\end{multline}
\begin{multline} \notag
 	i =1 :\: 2 p_0 ( \koff + \kon c \sqrt{\epsilon}) + p_{2} ( \kon c + \koff ) \\
 	= p_1 (2 \koff + \kon c (1+\epsilon) ),
 	\end{multline}
\begin{multline} \notag
 	i =0 :\: 2 p_0 ( \koff + \kon c \sqrt{\epsilon}) = p_1 (\kon c + \koff ). \hfill
\end{multline}
The factors of $\sqrt{\epsilon}$ arise when $i=0$ and there are two 
leading tips sharing the load, halving the force per filament. These 
equations lead to recurrence relations for $p_i$. Although $i$ cannot exceed
the length of the filament, the $p_i$ vanish exponentially as $i$ increases so we 
approximate $i_{max} = \infty$. Imposing normalization, $\sum p_i = 1$, we find
\begin{equation}
 	p^b_0 = \frac{c (1-\epsilon)}{c(1+2\sqrt\epsilon - \epsilon) + 2 c^b},
\end{equation}
where $c^b \equiv \kboff/\kbon$. An equivalent expression for $p^p_0$ applies for the 
pointed end. The average polymerization velocity for these stationary probabilities is
\begin{eqnarray}
 	\lambda^b &=& (2\kbon c \sqrt{\epsilon})\, p_0 
		+ (\kbon c \epsilon - \kboff) (1-p_0)\label{eq::lambdan2} \\
 	&=& (\kbon c \epsilon - \kboff) 
		+ (\kbon c (2 \sqrt{\epsilon} - \epsilon) + \kboff )\, p_0, \notag
\end{eqnarray}
and is a nontrivial function of both $c$ and $F$ (through 
$\epsilon$). To solve for the treadmilling concentration at a given force, 
we insert the barbed and pointed equivalent versions of 
Eq.~\ref{eq::lambdan2} into $\lambda^p = - \lambda^b$, which is a cubic 
polynomial in $c$, and we extract the one real, stable root.

\begin{figure}
\includegraphics[width=8.6cm]{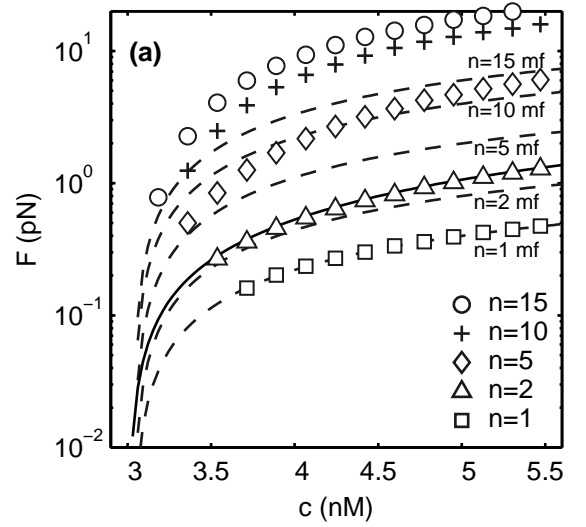}
\includegraphics[width=8.6cm]{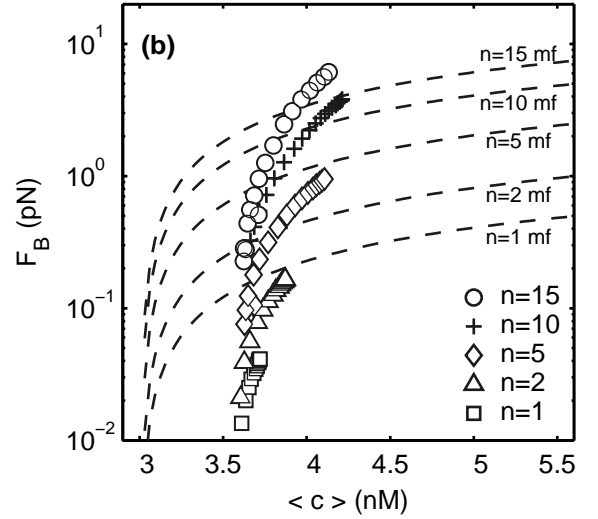}
\caption{Maximal steady-state force generation $F$ by a filament bundle with
a stochastic $n_{tip}$ as a function of steady-state concentration $c$. Various
average bundle thicknesses $n$ are shown. The 
dashed lines represent the mean-field predicted force-concentration 
relation (Eq.~\ref{eq::fstall}) if $n_{tip}=n$ independent tips were all sharing 
the load. (a) We hold $c$ fixed and allow an arbitrary bundle stiffness.
The solid line represents the analytic prediction for $n=2$ with a 
stochastic $n_{tip}$. The points indicate stochastic simulations, only allowing 
$n_{tip}$ to vary.  Fluctuations in $n_{tip}$ allow a significantly increased force
compared to the mean-field results. (b) The points indicate fully stochastic simulations, for the 
same $n$ as in a), and the average $c$ is used. A specific bundle elasticity is imposed by 
forcing $F=F_B$. Fluctuations in $c$ systematically decrease the bundle force compared to a), 
and this effect is stronger for smaller $c$.}
\label{fig::fvscss}
\end{figure}

\begin{figure}[h]
\includegraphics[width=4.1cm]{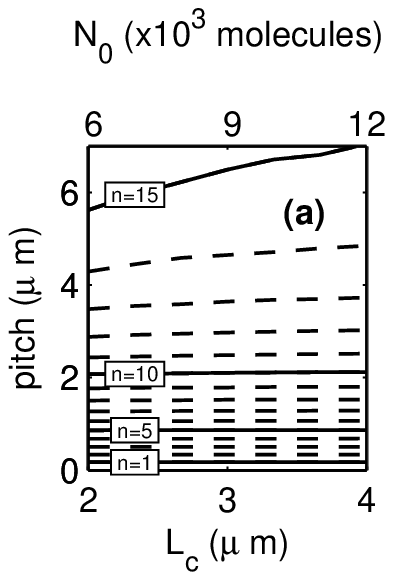}\includegraphics[width=4.3cm]{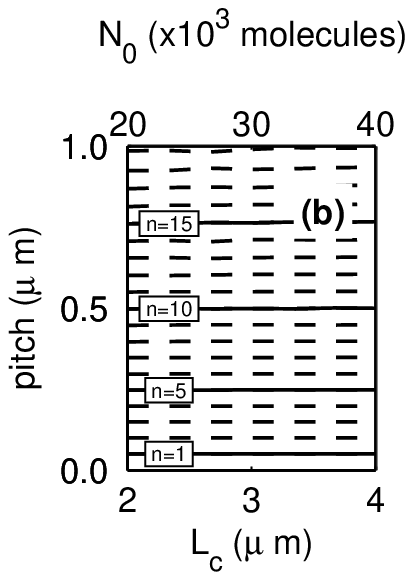}
\caption{Pitch $p$ as the cell length $L_c$ grows for various filament 
bulk thicknesses $n_{bulk}$ as predicted by stochastic simulation, for (a) 
$N_0 = 3\times 10^3 L_c$ so that an average $3\mum$ cell contains $9000$ 
MreB monomers, corresponding to regimes towards the low-$N_0$ end of 
Fig.~\ref{fig::pvsn0}, and (b) $N_0 = 10^4 L_c$, corresponding to the 
high-$N_0$ end. For most bundle thicknesses, the pitch is effectively constant as 
the cell elongates. However, for thicker bundles in the low-$N_0$ regime, 
the helical pitch exhibits a significant dependence on cell length, expanding 
as the cell doubles in size.}
\label{fig::growth}
\end{figure}

\begin{figure}[h] 
\includegraphics[width=8.6cm]{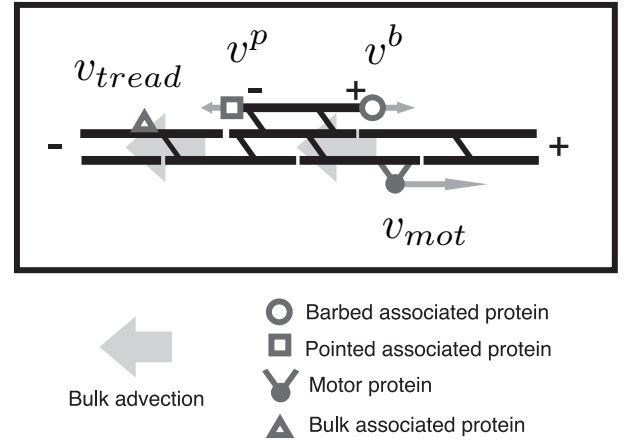}
\caption{Schematic of the different possible modes of transport along MreB bundles.  
Velocities are absolute with respect to the cell in the indicated directions. 
Bulk treadmilling advects bundles that span the cell length at a speed $v_{tread}$ 
towards the slow-growing pointed (``$-$'') end. Side associated protofilaments, without
constraint forces at their tips, have different polymerization rates and so are not
simply advected with the bundle. Their tip velocities are $v_p$ and $v_b$ for pointed
and barbed ends, respectively. As discussed in the text, $v_b$ is always opposite 
the treadmilling direction. Putative motor proteins would probably
be polarized and would have a characteristic speed $v_{mot}$.  The figure 
illustrates one polarized protofilament bundle, however it is possible that two
oppositely polarized bundles would exist within a cell --- in which case the polarity
and velocities of the second bundle should be opposite the first.}
\label{fig::translocationcartoon}
\end{figure}

\begin{figure}[h] 
\includegraphics[width=8.6cm]{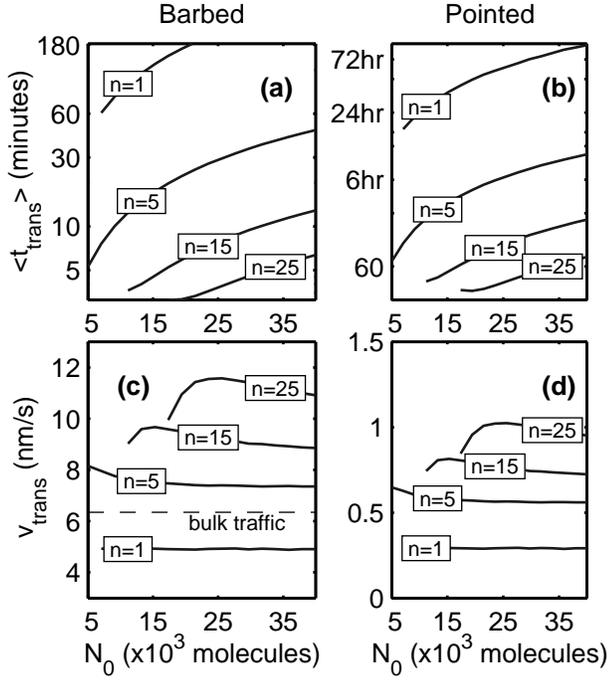}
\caption{Mean translocation times $\langle t_{trans}\rangle$ for 
macromolecule passengers being transported by MreB side protofilaments 
treadmilling on the side of the main cables vs. $N_0$. Constraint forces at the
ends of the helix keep $c$ above $c_c$, so that significant
axial movement can be seen with respect to bulk treadmilling. 
Different modes of transport along the cables are depicted schematically 
in Fig.~\ref{fig::translocationcartoon}. Barbed ends move antiparallel with respect to bulk 
advection (a,c). Pointed ends move in the same direction 
as bulk advection, however the slower polymerization rates lead to much longer 
translocation times (b,d).  Lateral traffic associated with parts of the 
bulk cables move towards the pointed end at $a_0 \lambda_{tread}$, indicated in 
c) by the dashed line.  Any passenger dissociation and re-association from protofilament tips 
will increase translocation times and decrease effective speeds.}
\label{fig::translocation}
\end{figure}

The relation between the total applied force, $F_B$, and the cytoplasmic monomer
concentration $c$ is shown by the solid line in Fig.~\ref{fig::fvscss}. Note 
that this derivation (and the simulations in Fig.~\ref{fig::fvscss}) 
imposes a static $c$, though $n_{tip}$ fluctuates. The disagreement
with the mean-field result (dashed line), indicates that the $n_{tip}$ fluctuations
are important for the total force.  

\small

\begin{thebibliography}{10}

\bibitem{mre52}
B. Alberts, D. Bray, J. Lewis, M. Raff and K. Roberts, {\em Molecular biology of the cell}, 4th ed. (Garland
  Science, New York, 2002).

\bibitem{mre86}
K.~A. Michie and J. L\H{o}we, Annu. Rev. Biochem. {\bf 75},  467  (2006).

\bibitem{mre55}
J.~M\o ller-Jensen, J. Borch, M. Dam, R. Jensen, P. Roepstorff and K. Gerdes, Mol. Cell {\bf 12},  1477  (2003).

\bibitem{mre92}
R. Carballido-Lopez, Microbiol. Mol. Biol. Rev. {\bf 70},  888  (2006).

\bibitem{mre87}
Y.-L. Shih and L. Rothfield, Microbiol. Mol. Biol. Rev. {\bf 70},  729  (2006).

\bibitem{mre30}
R.~A. Daniel and J. Errington, Cell {\bf 113},  767  (2003).

\bibitem{mre13}
L.~J.~F. Jones, R. Carballido-Lopez and J. Errington, Cell {\bf 104},
  913  (2001).

\bibitem{mre90}
H.~J. Defeu~Soufo and P.~L. Graumann, Mol. Micro. {\bf 62},  1340
  (2006).

\bibitem{mre73}
T. Kruse, J. M\o ller-Jensen, A. L\o bner-Olesen and K. Gerdes, EMBO
  {\bf 22},  5283  (2003).

\bibitem{mre71}
S.~Y. Kim, Z. Gitai, A. Kinkhabwala, L. Shapiro and W.~E. Moerner, Proc. Natl Acad. Sci. USA  {\bf 103}, 10929 (2006).

\bibitem{mre25}
H.~J.~D. Soufo and P.~L. Graumann, EMBO  {\bf 5},  789  (2004).

\bibitem{mre45}
Y.-L. Shih, T. Le and L. Rothfield, Proc. Natl Acad. Sci. USA {\bf 100},  7865
  (2003).

\bibitem{mre23}
R.~M. Figge, A.~V. Divakaruni and J.~W. Gober, Mol. Micro. {\bf
  51},  1321 (2004).

\bibitem{mre88}
R. Carballido-L{\'o}pez, A. Formstone, Y. Li, S.~D. Ehrlich, P. Noirot and J. Errington, Dev. Cell {\bf 11},  399
  (2006).

\bibitem{mre29}
R. Carballido-Lopez and J. Errington, Dev. Cell {\bf 4},  19  (2003).

\bibitem{mre33}
A. Fiebig and J.~A. Theriot, Proc. Natl Acad. Sci. USA {\bf 101},  8510  (2004).

\bibitem{protofilament}
{We use ``protofilament'' as a single polymerized strand of MreB monomers,
  ``bundle'' as one or more laterally associated protofilaments, while
  ``filament'' and ``cable'' are generic terms for similar structures.}

\bibitem{mre39}
H.~P. Erickson, Nature {\bf 413},  30  (2001).

\bibitem{mre67}
A. Formstone and J. Errington, Mol. Micro. {\bf 55},  1646  (2005).

\bibitem{mre65}
C.~W.~Wolgemuth, Y.~F.~Inclan, J.~Quan, S.~Mukherjee, G.~Oster and M.~A.~R.~Koehl, Phys. Biol. {\bf 2},  189  (2002).

\bibitem{mre37}
Y.-L. Shih, I. Kawagishi and L. Rothfield, Mol. Micro. {\bf 58},
  917  (2005).

\bibitem{mre34}
Z. Gitai, N. Dye and L. Shapiro, Proc. Natl Acad. Sci. USA  {\bf 101},  8643  (2004).

\bibitem{mre50}
T. Kruse, B. Blagoev, A. L\o bner-Olesen, M. Wachi, K. Sasaki, N. Iwai, M. Mann and K. Gerdes, Genes Dev. {\bf 20},  113  (2006).

\bibitem{mre53}
O. Espeli, P. Nurse, C. Levine, C. Lee and K. Marians, Mol. Micro. {\bf 50},  495  (2003).

\bibitem{mre26}
Z. Gitai, N.~A. Dye, A. Reisenauer, M. Wachi and L. Shapiro, Cell {\bf 120},  329  (2005).

\bibitem{blaauwen} A. Karczmarek, R.~M.-A. Baselga, S. Alexeeva, F.~G. Hansen, M. Vicente, N. Nanninga, T. den Blaauwen,
Mol. Micro. {\bf 65}, 51 (2007).

\bibitem{mre91}
C.~D. Webb, P.~L. Graumann, J.~A. Kahana, A.~A. Teleman, P.~A. Silver and R. Losick, Mol. Micro. {\bf 28},  883  (1998).

\bibitem{act04}
A. Mogilner and G. Oster, Curr. Biol. {\bf 13},  721  (2003).

\bibitem{act05}
T.~D. Pollard and G.~G. Borisy, Cell {\bf 112},  453  (2003).

\bibitem{act01}
A. Mogilner and G. Oster, Biophys. J. {\bf 71},  3030  (1996).

\bibitem{mre38}
O. Esue, D. Wirtz and Y. Tseng, J. Bacteriol. {\bf 188},  968
  (2006).

\bibitem{mre28}
D.-J. Scheffers, L.~J.~F. Jones,and J. Errington, Mol. Micro. {\bf
  51},  749  (2004).

\bibitem{mre72}
F. van~den Ent, L. Amos and J. L{\"o}we, Nature {\bf 413},  39  (2001).

\bibitem{act20}
A. Mogilner and L. Edelstein-Keshet, Biophys. J. {\bf 83},  1237
  (2002).

\bibitem{act19}
L. Edelstein-Keshet and G. Ermentrout, J. Math. Biol. {\bf
  40},  64  (2000).

\bibitem{mre98}
G.~I. Bell, Science {\bf 200},  618  (1978).

\bibitem{act23}
T.~S. Karpova, J.~G. McNally, S.~L. Moltz and J.~A. Cooper, J. Cell Biol. {\bf
  142},  1501  (1998).

\bibitem{mre07}
T. Kruse, J. Bork-Jensen and K. Gerdes, Mol. Micro. {\bf 1},
  78  (2005).

\bibitem{ela14}
N. Chouaieb, A. Goriely and J.~H. Maddocks, Proc. Natl Acad. Sci. USA {\bf 103},  9398  (2006).

\bibitem{ela01}
S.~S. Antman, {\em Nonlinear Problems of Elasticity}, 2nd ed. (Springer,
  New York, 2004).

\bibitem{ela16}
C. Heussinger, M. Bathe and E. Frey, cond-mat/0702097.

\bibitem{mre89}
J. Stricker, P. Maddox, E.~D. Salmon and H.~P. Erickson, Proc. Natl Acad. Sci. USA {\bf 99},  3171
  (2002).

\bibitem{act25}
M. {Rief}, R.~S. Rock, A.~D. Mehta, M.~S. Mooseker, R.~E. Cheney and J.~A. Spudich, Proc. Natl Acad. Sci. {\bf
  97},  9482  (2000).

\bibitem{act28}
A.~L. Wells, A.~W. Lin, L.-Q. Chen, D. Safer, S.~M. Cain, T. Hasson, B.~O. Carragher, R.~A. Milligan and H.~L. Sweeney, Nature {\bf 401},  505  (1999).

\bibitem{act27}
D. Kovar and T. Pollard, Nature Cell Biology {\bf 6},  1158  (2004).

\bibitem{mre27}
E.~C. Garner, C.~S. Campbell and R.~D. Mullins, Science {\bf 306},  1021
  (2004).

\bibitem{act21}
L. Edelstein-Keshet and G. Ermentrout, Bull. Math. Biol. {\bf
  60},  449  (1998).

\bibitem{act29}
E. Evans, D. Berk and A. Leung, Biophys. J. {\bf 59},  838  (1991).

\bibitem{mre48}
H. Lam, W.~B. Schofield and C. Jacobs-Wagner, Cell {\bf 124},  1011  (2006).

\bibitem{mre93}
W. Houry, D. Frishman, C. Eckerskorn and F. Lottspeich, Nature {\bf 402},  147
   (1999).

\bibitem{mre96}
H. Ogino, M. Wachi, A. Ishii, N. Iwai, T. Nishida, S. Yamada, K. Nagai and M. Sugai, Genes to Cells {\bf 9},  765  (2004).

\bibitem{mre94}
J.~E. Cabrera and D.~J. Jin, Mol. Micro. {\bf 50},  1493  (2003).

\bibitem{mre95}
J.~E. Cabrera and D.~J. Jin, J. Bacteriol. {\bf 188},  4007  (2006).

\bibitem{mre104}
S.~S.~Andrews and A.~P.~Arkin, Biophys. J. BioFAST 106.102343  (2007).

\bibitem{ela15}
A. Levine, T. Liverpool and F. MacKintosh, Phys. Rev. Lett. {\bf 93},
  38102  (2004).

\bibitem{ela13}
G. van~der Heijden and J. Thompson, Nonlinear Dynamics {\bf 21},  71  (2000).

\bibitem{mre49}
X. Yao, M. Jericho, D. Pink and T. Beveridge, J. Bacteriol. {\bf 181},  6865
  (1999).

\bibitem{act30}
J. Howard, {\em Mechanics of Motor Proteins and the Cytoskeleton} (Sinauer Associates, Sunderland, MA, 2001)

\bibitem{act03}
A. Mogilner and G. Oster, Biophys. J. {\bf 84},  1591  (2003).

\bibitem{act07}
X. Liu and G.~H. Pollack, Biophys. J. {\bf 83},  2705  (2002).

\bibitem{act26}
K. Luby-Phelps, D. Taylor and F. Lanni, J. Cell Biol. {\bf 102},  2015
  (1986).

\end{thebibliography}

\begin{table*} {\small 
\begin{tabular}{llll}
\hline
  Symbol & Value & Meaning & Reference\\
\hline
\hline
\multicolumn{4}{l}{Eukaryotic actin kinetic rates for polymerization} \\
$\kbon$ & $12 \muM^{-1} \s^{-1}$ & Barbed-end addition rate constant & \citel{act05} \\
$\kpon$ & $1.3 \muM^{-1} \s^{-1}$ & Pointed-end addition rate constant & \citel{act05} \\
$\kboff$ & $1.4 \s^{-1}$ & Barbed-end dissociation rate & \citel{act05} \\
$\kpoff$ & $0.8 \s^{-1}$ & Pointed-end dissociation rate  & \citel{act05} \\
$c_c$ & $0.167 \muM$ &critical concentration & (Eq.~\ref{eq::ccdef} and \citel{act05})\\
$\lambda_{tread}$ & $0.58 \s^{-1}$ & free treadmilling rate & (Eq.~\ref{eq::lambdatreaddef} and \citel{act05})\\
\hline
\multicolumn{4}{l}{Assumed MreB kinetic rates for polymerization} \\
$\kbon$ & $1360 \muM^{-1} \s^{-1}$ & Barbed-end addition rate constant & (scaled to fit $c_c$ and $\lambda$) \\
$\kpon$ & $150 \muM^{-1} \s^{-1}$ & Pointed-end addition rate constant & (scaled to fit $c_c$ and $\lambda$) \\
$\kboff$ & $2.9 \s^{-1}$ & Barbed-end dissociation rate & (scaled to fit $c_c$ and $\lambda$)\\
$\kpoff$ & $1.7 \s^{-1}$ & Pointed-end dissociation rate  &(scaled to fit $c_c$ and $\lambda$)\\
$c_c$ & $0.003 \muM$ &critical concentration & \citel{mre38}\\
$\lambda_{tread}$ & $1.2 \s^{-1}$ & free treadmilling rate & \citel{mre71}\\
\hline
$B$ & $1.0 \times 10^4 \pN \nm^2$ & Bending modulus of MreB protofilament &  (from actin, \citel{act07}) \\
$E$ & $ 20 \pN$ & Stretching modulus of MreB protofilament & (from actin, \citel{act07})\\ 
$a_0$ & $ 5.1 \nm$ & Monomer spacing of MreB &  \citel{mre38}\\ 
$R_c$ & $400 \nm$ & Radius of \bsub\ cell & \citel{mre88}\\
$L_c$ & $(2 - 4) \times 10^3 \nm$ & Length of cylindrical portion of \bsub\ cell & \citel{mre88}\\
$\eta$ & $10^{-9} \pN s / \nm^2 $ & Cytoplasmic viscosity & \citel{act26}\\
\hline\label{table::paramtable} \end{tabular}
\caption{The parameters used in this paper for MreB and its paralogues. Only $c_c$ and $a_0$ have been 
directly measured for MreB from {\em T. maritima}. 
Other assumed parameters have been extracted from the properties of 
eukaryotic actin, which is homologous. The kinetic rates chosen for MreB recover $c_c$ \citel{mre38} and 
the \invivo\ treadmilling rate \citel{mre71}, but leave the pointed/barbed asymmetry unchanged. The 
elastic moduli are for a single protofilament of F-actin ($n=1$).}} 
\end{table*}

\end{document}